# End-to-end joint optimization of metasurface and image processing for compact snapshot hyperspectral imaging

Qiangbo Zhang [a], Zeqing Yu [a], Xinyu Liu [a], Chang Wang [a,b,*], Zhenrong Zheng [a,b,*]

[a] *State Key Laboratory of Modern Optical Instrumentation, College of Optical Science and Engineering, Zhejiang University, Hangzhou 310027, China*
[b] *Intelligent Optics & Photonics Research Center, Jiaxing Research Institute Zhejiang University, Jiaxing 314000, China*
* changwang_optics@zju.edu.cn
* zzr@zju.edu.cn

**Abstract:** Traditional snapshot hyperspectral imaging systems generally require multiple refractive-optics-based elements to modulate light, resulting in bulky framework. In pursuit of a more compact form factor, a metasurface-based snapshot hyperspectral imaging system, which achieves joint optimization of metasurface and image processing, is proposed in this paper. The unprecedented light manipulation capabilities of metasurfaces are used in conjunction with neural networks to encode and decode light fields for better hyperspectral imaging. Specifically, the extremely strong dispersion of metasurfaces is exploited to distinguish spectral information, and a neural network based on spectral priors is applied for hyperspectral image reconstruction. By constructing a fully differentiable model of metasurface-based hyperspectral imaging, the front-end metasurface phase distribution and the back-end recovery network parameters can be jointly optimized. This method achieves high-quality hyperspectral reconstruction results numerically, outperforming separation optimization methods. The proposed system holds great potential for miniaturization and portability of hyperspectral imaging systems.

## 1. Introduction

Hyperspectral imaging captures the spectral distribution of a scene as a data cube that describes the spectral intensity of each wavelength at each pixel location. At present, hyperspectral imaging technology has been widely applied in many fields, such as agriculture surveillance [1], food safety [2], face recognition [3], military reconnaissance [4], etc.

For acquiring the huge spectral data cube, traditional spectral scanning systems require long exposure time which limits their use in real-time applications [5,6], thus a number of snapshot spectral imaging systems consisting of cascaded refractive-optics-based devices have been proposed for faster imaging. However, the improved functionality comes at the expense of bulky volume, impeding their portability [7,8]. To solve this problem, diffractive optical elements (DOEs) are introduced to snapshot hyperspectral imaging systems, facilitating the system compactness and portability [9–11]. Especially, Jeon et al. achieved high-quality spectral image reconstruction by applying a single DOE in front of the bare sensor [9]. Yet, the defect of only modulating the phase of light limits the further development of DOE-based hyperspectral imaging systems.

Metasurfaces, composed of two-dimensional arrays of subwavelength optical scatterers, are regarded as powerful substitutes to conventional diffractive and refractive optics [12–18]. Compared with DOEs, metasurfaces inherently has a higher space-bandwidth product due to its subwavelength pitch [19]. In addition, metasurfaces with more powerful wavefront manipulation capabilities can steer the phase, amplitude, and polarization of light, while DOEs merely control the phase of light [17,18]. Recently, the combination of metasurfaces and computational imaging has yielded remarkable results [19–25], such as full-color achromatic imaging [19], three-dimensional imaging [23], etc. However, the end-to-end snapshot hyperspectral imaging system based on joint optimization of the metasurface and image processing has not been well investigated.

In this paper, a compact snapshot hyperspectral imaging system implemented by joint optimization of metasurface and image processing is proposed. It takes advantage of the extremely strong dispersion of metasurfaces to construct the point spread function (PSF) that varies greatly with wavelengths and provides extra freedom for hyperspectral imaging. Simultaneously, an end-to-end neural network based on joint optimization of metasurface phase distribution and image decoding is developed for hyperspectral imaging reconstruction. Fig. 1 shows the overview of this hyperspectral image reconstruction system. The metasurface-based image formation is modeled fully differentiable by consisting of the following differentiable steps: metasurface phase design, PSF calculation, convolution operation, sensor capture, and noise addition. Hyperspectral images can be reconstructed by the prior-based unfolding recovery network and then compared with the ground true to calculate the loss function for backward propagation. During backward propagation, the recovery network parameters and metasurface phase distribution as optimizable variables are updated. Compared with manually setting the phase distribution, our proposed hyperspectral imaging system can achieve better reconstruction results through the co-optimization of front-end optics and back-end recovery networks. In summary, this proposed hyperspectral imaging system realizes high image recovery performance with a compact form factor, further promoting the miniaturization and portability of hyperspectral imaging systems.

Specifically, the contributions of the proposed system are summarized as follow:

1. We introduce the geometric metasurface into the compact snapshot hyperspectral imaging system, exploiting its strong dispersion to distinguish spectral information.

2. We achieve joint optimization of hardware and recovery algorithm by differentiating the phase distribution of the geometric metasurface, enabling it to be co-optimized with the hyperspectral reconstruction network for optimal performance.

3. A back-end prior-based unfolding recovery network is employed as the hyperspectral reconstruction network. It is developed by joint optimization with the front-end metasurface, achieving better hyperspectral reconstruction results than those of separation optimization methods.

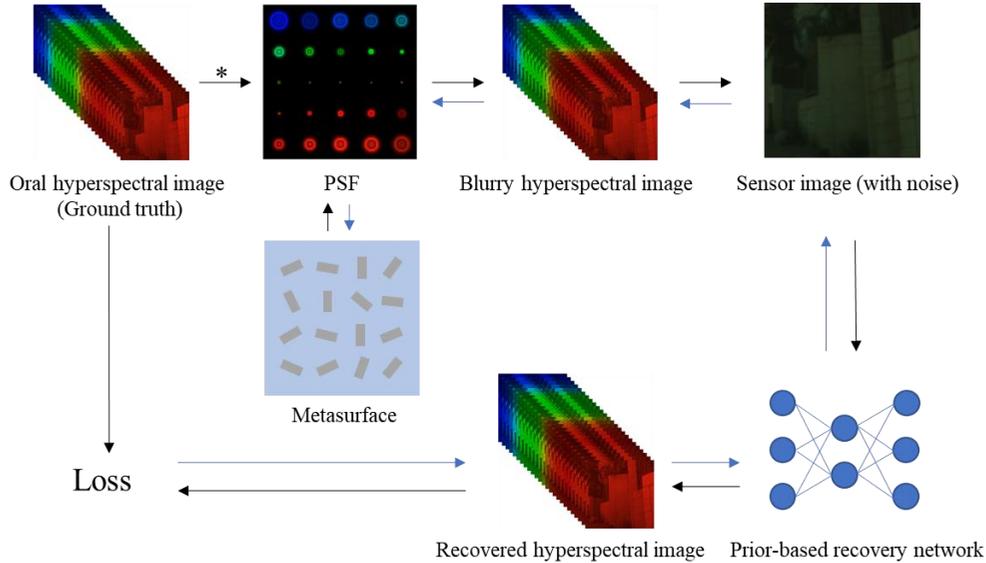

Fig. 1. Overview of the proposed hyperspectral image reconstruction system. Black arrows and blue arrows indicate forward propagation and back propagation, respectively.

## 2. Method

### 2.1 Metasurface design

Phase profile is a crucial part of metasurfaces design. For geometric metasurfaces, phase modulation can be ideally defined as $J(x, y)|L\rangle=e^{-i2\theta} |R\rangle$, where $J(x, y)$ and $\theta$ are respectively the Jones matrix and rotation angle of each unit nanopillar, $|L\rangle$ and $|R\rangle$ denote left-circularly polarized (LCP) and right-circularly polarized (RCP) light. When the geometry of unit nanopillar is constant and LCP light incidents on a geometric metasurface, the phase change of the unit cell is only relevant to the rotation angle of the nanopillar and independent of the incident wavelengths. Therefore, geometric metasurfaces can simplify phase profile design by maintaining phase distribution for broad bandwidth.

The schematic diagram of the unit cell is depicted in Fig. 2(a), where silicon dioxide ($SiO_2$) and silicon nitride ($Si_3N_4$) are chosen as the substrate and the nanopillar, respectively. Full-wave finite-difference time-domain (FDTD) simulations are employed to optimize silicon nitride nanofins of different structural parameters for obtaining peak polarization conversion efficiency (PCE) from LCP to RCP in the 460-700 nm band, which is shown in Fig. 2(b). Finally, the chosen unit cell parameters are $L = 315$ nm, $W = 115$ nm, $H = 750$ nm and $C = 400$ nm.

The metasurface phase profile $\phi$ with respect to current polar coordinate $r$ is designed in the form of

$$\phi(r) = \frac{2\pi}{\lambda}\left(f - \sqrt{r^2 + f^2}\right) + \sum_{i=0}^{n} a_i \left(\frac{r}{R}\right)^{2i}, \quad (1)$$

where the first term is defined as the hyperboloidal phase $\phi_1(r)$, other terms denote the polynomial phase $\phi_2(r)$, $\lambda$ means the default wavelength, $f$ is the focal length, $R$ is the radius of the metasurface, $\{a_0, …a_n\}$ indicate optimizable coefficients and $n$ is the number of polynomial terms. Compared with phase distributions of general metalenses [13,14], this designed phase distribution is further optimized by adding polynomial phase on the basis of the hyperboloidal phase. By optimizing polynomial coefficients $\{a_0, …a_n\}$, the phase distribution can be changed more flexibly to achieve better spectral imaging quality.

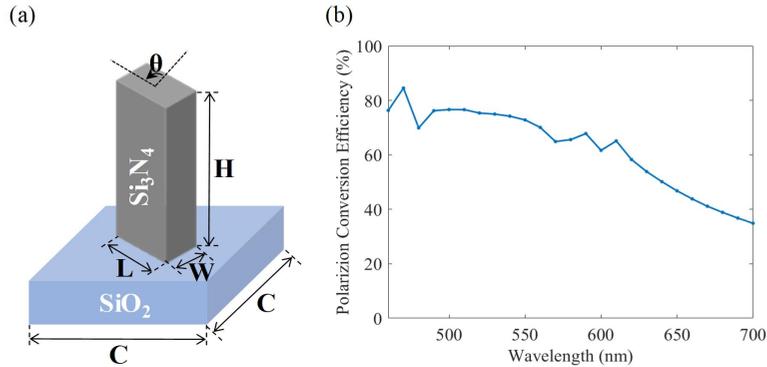

Fig. 2. (a) Schematic of the unit cell containing a silicon nitride nanopillar on a silica substrate. (b) Polarization conversion efficiency from LCP to RCP of the unit cell in the 460~700nm band.

### 2.2 PSF calculate

In this section, the PSF of the metasurface is calculated by introducing the model of diffraction imaging and the schematic diagram of the diffraction imaging is demonstrated in Fig. 3. The incident light passes through the left-circularly polarizer, the metasurface, and the right-circularly polarizer successively to be received by the sensor. Suppose the light source is located at infinity, the parallel monochromatic light of wavelength $\lambda$ with normalized amplitude passes through the left-handed circular polarizer and then vertically hits on the metasurface. The wave field $u_{meta}$ at position $(x', y')$ along the metasurface after passing through the metasurface can be formulated as

$$u_{meta}(x',y') = \sqrt{\frac{1}{2}PCE_\lambda} \exp\left[i\left(\phi_{meta}(x',y')\right)\right], \qquad (2)$$

where $PCE_\lambda$ is polarization conversion efficiency of the metasurface for different wavelengths, and $1/2$ indicates the energy loss caused by the polarizer. The sensor is located at the distance of $z$ behind the metasurface. When $z \gg \lambda$, the wave field $u_{sensor}(x, y)$ can be calculated approximated by the Fresnel diffraction as

$$u_{sensor}(x,y) = \frac{1}{i\lambda z}\exp(ikz)\iint u_{meta}(x',y')\exp\left\{\frac{ik}{2z}\left[(x-x')^2+(y-y')^2\right]\right\}dx'dy', \qquad (3)$$

where $k = 2\pi / \lambda$ is the wavenumber. Finally, the PSF becomes the distribution of light intensity, that is, the square value of the wave field $u_{sensor}$. The PSF $P_\lambda(x, y)$ can be derived by Fourier transform as:

$$P_\lambda(x,y) \propto \left|\mathcal{F}\left\{\sqrt{\frac{1}{2}PCE_\lambda}\exp\left\{i\left[\phi_{meta}(x',y')+\frac{\pi}{\lambda z}(x'^2+y'^2)\right]\right\}\right\}\right|^2. \qquad (4)$$

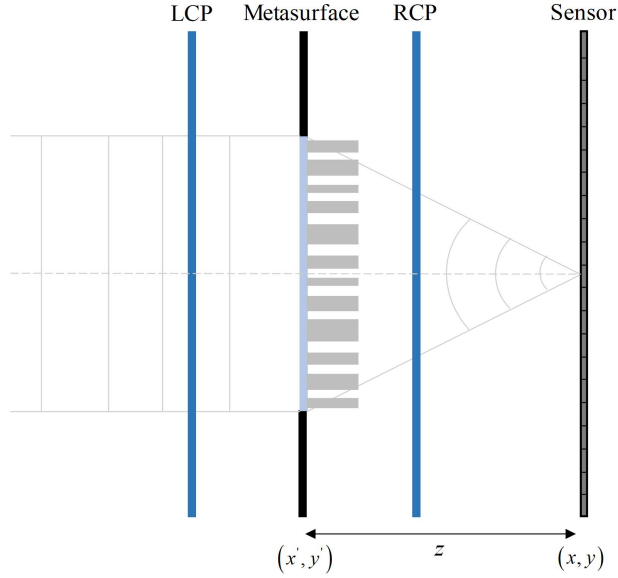

Fig. 3. Schematic diagram of diffraction imaging with the metasurface

*2.3 Sensor image formation*

Since the hyperspectral images of $c$ channels are captured by a traditional RGB image sensor, it is essential to apply the spectral response of the sensor to compress the $c$ spectral channels into three RGB channels. Meanwhile, considering the noise $n$ during imaging, the final RGB image $J_{RGB}$ received from the sensor can be expressed as

$$J_{RGB} = \int (I_\lambda * P_\lambda) \cdot S_{\lambda,rgb} d\lambda + n, \quad (5)$$

where $I_\lambda$ denotes the oral hyperspectral image, $S_{\lambda,rgb}$ is the spectral sensitivity function of the sensor and $*$ is defined as the convolution operator. Further, Eq. can be rewritten in the form of vector and matrix:

$$J = \Phi I + N, \quad (6)$$

where $J \in \mathbb{R}^{HW3 \times 1}$ and $I \in \mathbb{R}^{HWc \times 1}$ are respectively defined as the captured image vector and oral hyperspectral image vector, $H$ and $W$ are the image size. $\Phi = SP$, where $S \in \mathbb{R}^{HW3 \times HWc}$ and $P \in \mathbb{R}^{HWc \times HWc}$ denote the sensor spectral sensitivity matrix and PSF matrix, respectively. $N \in \mathbb{R}^{HW3 \times 1}$ represents the noise vector.

### 2.4 Optimization problem

Considering the fact that the process of hyperspectral image reconstruction is from blurred to clear, from noisy to denoised, and from RGB three-channel to hyperspectral multi-channel, hyperspectral reconstruction requires a prior item since it is a very underdetermined problem. The unfolding neural network based on image prior has proved to be the very helpful non-blind convolution solver for such inverse image restoration problems[9,26–28], but has not yet been jointly optimized with the front-end optics. Therefore, we also employ the prior-based unfolding neural network as the back-end recovery algorithm jointly optimized with the metasurface. The hyperspectral reconstruction objective function including the spectral prior can be expressed as

$$I = \arg\min_I \frac{1}{2} \|J - \Phi I\|_2^2 + V(I), \quad (7)$$

where the first item is defined as the spectral data item, and the second item indicates the spectral prior item, with $V(\cdot)$ denoting the prior function of hyperspectral images. Using the half-quadratic splitting method, the data term and prior term of Eq. can be decoupled and transformed into an unconstrained optimization problem by introducing auxiliary variables $H \in \mathbb{R}^{HW3 \times 1}$:

$$(I, H) = \arg\min_{I,H} \frac{1}{2} \|J - \Phi I\|_2^2 + \eta \|I - H\|_2^2 + V(H) \quad s.t. \quad H = I, \quad (8)$$

where $\eta$ is the penalty parameter. Further, Eq. can be solved by alternately solving the two subproblems:

$$I^{(t+1)} = \arg\min_I \|J - \Phi I\|_2^2 + \eta \|I - H^{(t)}\|_2^2, \quad (9)$$

and

$$H^{(t+1)} = \arg\min_H \eta \|I^{(t+1)} - H\|_2^2 + V(H). \quad (10)$$

An inexact solution of $I^{(t+1)}$ can be calculated by alternating gradient descent:

$$\begin{aligned} I^{(t+1)} &= I^{(t)} - \delta \left[ \Phi^T (\Phi I^{(t)} - J) + \eta (I^{(t)} - H^{(t)}) \right] \\ &= \overline{\Phi} I^{(t)} + \delta I^{(0)} + \delta \eta H^{(t)}, \end{aligned} \quad (11)$$

where $\overline{\Phi} = [(1-\delta\eta)I - \delta\Phi^T\Phi]$, $I^{(0)} = \Phi^T J$, and $\delta$ is the parameter controlling the gradient descent step size. Now, $H^{(t)} = R(I^{(t)})$ can be firstly solved through a neural network of spectral prior, where $R(\cdot)$ is the nonlinear function from $I^{(t)}$ to $H^{(t)}$. Then $I^{(t+1)}$ can be solved by Eq. . After iteratively solving $H^{(t)}$ and $I^{(t)}$, the optimum spectral image $I^{(T)}$ can eventually be recovered.

### 2.5 Network architecture

The network framework shown in Fig. 4(a) is utilized to iteratively calculate $I^{(T)}$, with the RGB image $J$ received by the sensor as the only input, and the output of the first iteration $I^{(1)}$ is obtained according to the Eq. . By repeating the same iteration step continuously, finally the output $I^{(T)}$ is generated. In the network framework, prior network is the nonlinear function from $I^{(t)}$ to $H^{(t)}$, and the classical U-net network [29] is adopted to calculate the spectral prior. As shown in Fig. 4(b), the structure of the prior network consists of a contracting path and a symmetric expansive path. The input $I^{(t)}$ is first extracted to 64 feature channels by two convolutional layers, entering to the contracting path where each block contains a max-pooling layer for downsampling and two convolutional layers for extracting features. Next in the extensive path, the upsampling block and the corresponding block in the contracting path are concatenated together as the input of two convolutional layers. Finally, a convolutional layer is performed on the output of the extensive path to obtain the output $H^{(t)}$ of the prior network. In this paper, the number of iterations $T$ is set to be 6. Meanwhile, in order to increase the flexibility of the network, $\delta$ and $\eta$ are set as optimizable variables and are different at each stage [9,26].

The loss function is defined as the sum of the mean squared errors between each intermediate result and the ground truth:

$$Loss = \sum_{t=1}^{T} \left\| I^{(t)} - I \right\|_2^2. \tag{12}$$

This recursive supervised approach is able to accelerate training and the learned optimizer steadily pushes each intermediate result to be closer to the ground truth [28].

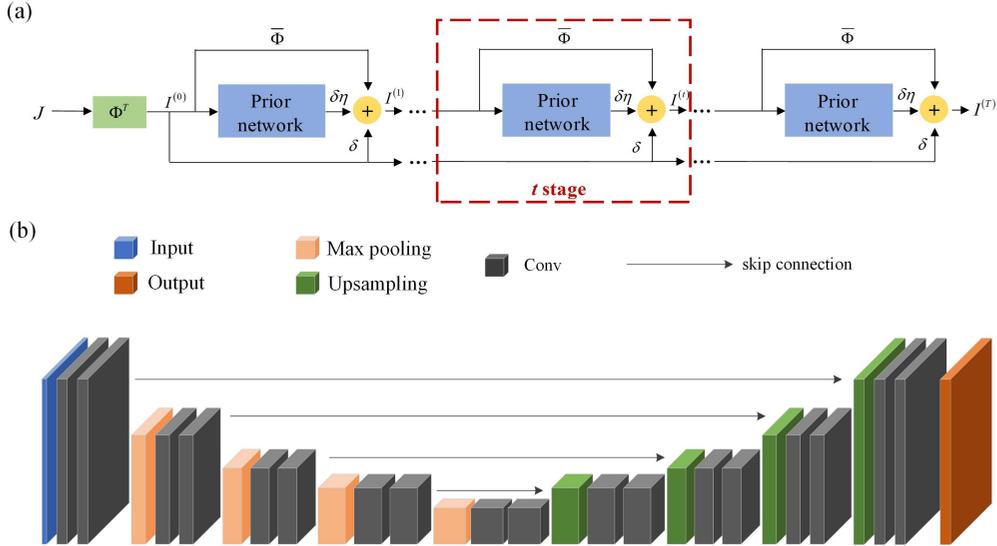

Fig. 4. The network architecture of our method for hyperspectral image restoration. (a) The overall architecture of the proposed deep neural network. (b) The architecture of the hyperspectral image prior network.

## 3. Results and discussions

### 3.1 Implementation details

In this paper, the joint optimization of front-end metasurface and back-end recovery network is utilized to restore hyperspectral images, with the polynomial phase coefficients and recovery network parameters being the optimizable variables. Specifically, metasurface phase distribution is firstly set to be the hyperboloidal phase $\phi_1$ and then only the back-end recovery network parameters are optimized. After the loss function converges, the polynomial phase $\phi_2$ is added and jointly optimized with the parameters of the recovery network. In addition, the alternate update method is employed to optimize the phase coefficients and recovery network parameters for 10 and 30 iterations, respectively [19].

During training, the publicly available ICVL dataset [30] (150 training and 22 testing images) is utilized. The training hyperspectral images include 25 wavelength channels in a range from 460 nm to 700 nm are randomly cropped to 128 × 128 as the input of the entire hyperspectral imaging system. For metasurface design, the focal length is designed as $f = 15mm$, the diameter is 1 mm, the default wavelength is set as 580 nm, the number of polynomial terms $n = 8$. The spectral response of the Nikon D700 camera is applied to compress hyperspectral channels into RGB channels, and the pixel size of the sensor is set to be 5 $\mu$m. The distance between the metasurface and the sensor is set to be 15 mm. In the sensor image formation model, gaussian noise is added with standard deviation of $\sigma = 0.005$. The coefficients of polynomial phase and network parameters are optimized by ADAM optimizer ($\beta_1 = 0.9$, $\beta_2 = 0.999$). The learning rates are all set to be $10^{-4}$, and the batch size is set as 4. The equipment used includes an Intel Core i9-10900K CPU with 64GB memory and an NVIDIA 3090 GPU with 24GB memory. It takes 1.28s to restore a hyperspectral image with 800 × 800 resolution applying the proposed recovery network.

*3.2 Results*

Fig. 5 shows images and PSFs obtained at different stages during spectral reconstruction with the proposed method. The entire 3D hyperspectral data are synthesized into an RGB image for ease of display in this paper. Fig. 5(a) illustrates the oral hyperspectral image. The PSFs of 25 channels after optimization in the bandwidth from 460 nm to 700 nm are demonstrated in Fig. 5(b). The PSF is smallest when the wavelength is around 580nm and expands with the wavelength varying. Notably, PSFs impart additional spectral information to hyperspectral imaging by virtue of their wavelength-dependent feature. Fig. 5(c) exhibits the image received by the sensor, and it is dim due to the polarization-dependent energy loss. Meanwhile, large PSF kernels and noise lead to the blurry image and low signal-to-noise ratio (SNR). Fig. 5(d and e) show respectively the restored hyperspectral image and the reconstructed images for 25 spectral channels from 460 nm to 700 nm.

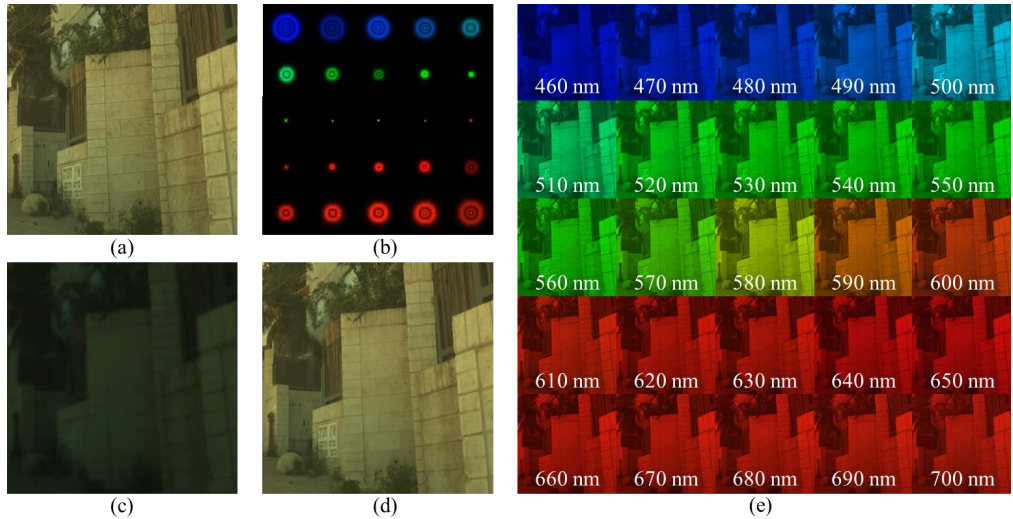

Fig. 5. (a) The oral hyperspectral image. (b) Simulated PSFs of 25 channels in the bandwidth from 460 nm to 700 nm. The wavelength channel sequence is the same as (e). (c) The color image received by the sensor. (d) The recovered hyperspectral image. (e) Recovered images for 25 spectral channels from 460 nm to 700 nm.

To further evaluate the proposed system, we retrain Jeon's rotative PSF method [9] (designing phase to make PSF rotate with wavelength varying, and we apply this to the metasurface) with our recovery network for comparison. Moreover, the phase design methods of only polynomial or hyperboloid are also added to the comparison with same recovery network. Specially, the phase of Jeon's and Hyperboloid methods are pre-configured and only the parameters of the recovery network are optimized, while the phase of Polynomial and Our methods are jointly optimized with the back-end recovery network.

Three quantitative image quality metrics are employed to evaluate the performance of these methods: average peak signal-to-noise ratio (PSNR), structural similarity (SSIM) [31], and spectral angle mapping (SAM) [32]. The performance of different methods is listed in Table 1. It can be seen that our method demonstrates the highest PSNR, and is nearly equally well in SSIM and SAM with the Hyperboloid method.

Table 1. Average PSNR, SSIM, and SAM comparison of four spectral reconstruction methods with the same test dataset. Bold text indicates the highest accuracy.

| Method | Jeon's | Polynomial | Hyperboloid | Ours |
|---|---|---|---|---|
| **PSNR(dB)** | 33.14 | 34.53 | 36.06 | **37.24** |
| **SSIM** | 0.962 | 0.974 | 0.980 | **0.981** |
| **SAM** | 0.048 | 0.058 | **0.039** | 0.040 |

To show more intuitive results, two representative sets of restored images are visualized as shown in Fig. 6. The results recovered by the proposed method not only achieve the best results on image quality metrics, but also have sharper edges in visual perception, outperforming those of other methods.

Additionally, Fig. 7 shows the recovered spectrum by different methods for two points in the selected image. Comprehensively, the reconstructed spectra using the proposed method is closest to the value of ground truth, followed by results of the Hyperboloid method. The basically close performance of these two methods is also consistent with the SAM data given in Table 1.

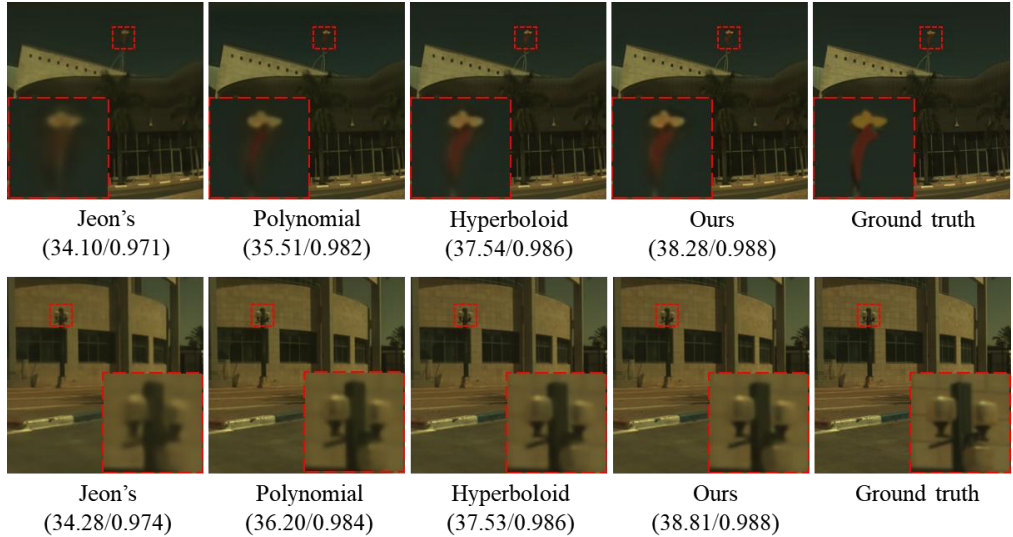

Jeon's (34.10/0.971)   Polynomial (35.51/0.982)   Hyperboloid (37.54/0.986)   Ours (38.28/0.988)   Ground truth

Jeon's (34.28/0.974)   Polynomial (36.20/0.984)   Hyperboloid (37.53/0.986)   Ours (38.81/0.988)   Ground truth

Fig. 6. Visual quality comparison of five methods. The PSNR and SSIM for the recovered results are shown in the parenthesis.

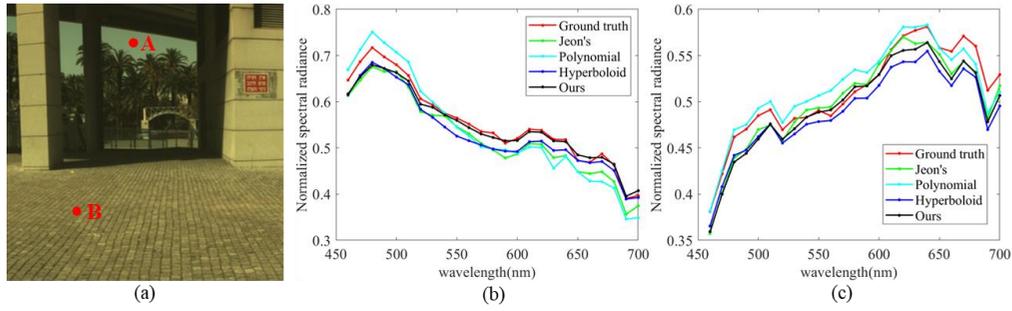

Fig. 7. (a) A hyperspectral image with two randomly selected spatial points A and B. (b and c) Respectively the reconstruction spectra of points A and B using different methods. The reconstruction spectra of this proposed method (the black line) is closest to the ground truth (red line)

### *3.3 Discussions*

A compact snapshot hyperspectral imaging system jointly optimized by metasurface phase and recovery network is here proposed. Since the optimizable polynomial phase is added to the basis of the general hyperboloidal phase, the phase distribution of the metasurface holds greater flexibility. Furthermore, the phase-adaptive learning to optimize polynomial coefficients is more efficient than the hand-designed phase, and the former can achieve better recovery quality.

However, there are still some limitations in this study. When designing the metasurface, it is defaulted that the rotation has no effect on the spectral response. But in fact, the spectral response of the unit cell changes slightly with the nanopillar rotating. However, considering that there will also be errors in the actual manufacturing process, an experimental calibration of PSF is ultimately required. After the PSF experimental calibration, the parameters of the recovery network need to be fine-tuned, after which high-quality spectral reconstruction results can be obtained.

During the PSF calculation, it is assumed that the light source is located at infinity. When the light source is at a finite distance in front of the metasurface, the PSF is not only a function of the spectrum, but also a function of distance and it has been explored by Baek et al [33]. In this paper, the PSF calculation of limited distance is not considered, making the proposed spectral reconstruction method suitable for shooting distant objects.

Although metasurfaces with an unparalleled ability can control the phase, amplitude, and polarization of light synchronously, this paper only exploits the phase modulation capabilities. Incorporating amplitude and polarization modulation into computational imaging should be put into effect in the future.

### 4. Conclusion

In this paper, we propose a compact and portable snapshot hyperspectral imaging system, which utilizes a metasurface and achieves joint optimization of hardware and recovery algorithm. The designed metasurface phase profile, consisting of fixed hyperboloidal phase and optimizable polynomial phase, is jointly optimized with the recovery network to restore the high-quality hyperspectral images, which outperform those of the separation optimization, and there may be opportunities to further improve by increasing the manipulation dimension of the metasurface. Taking advantage of the thin but powerful light manipulation ability of metasurfaces, hyperspectral imaging systems are expected to further realize the miniaturization and portability while maintaining high performance.

**Declaration of competing interest**

The authors declare that they have no known competing financial interests or personal relationships that could have appeared to influence the work reported in this paper.

**Data availability**

Data will be made available on request.

**Acknowledgement**

This work was supported by the Key R&D Plan of Zhejiang Province a under Grant No. 2022C01127.